\documentclass[aps,prd,twocolumn,a4paper,superscriptaddress,floatfix]{revtex4-1}
\usepackage{graphicx}
\usepackage{multirow}
\usepackage{latexsym}
\usepackage{hyperref}
\usepackage[british]{babel}
\usepackage{amsmath}

\begin{document}

\title{Astrophysical and local constraints on string theory: runaway dilaton models}
\author{C. J. A. P. Martins}
\email{Carlos.Martins@astro.up.pt}
\affiliation{Centro de Astrof\'{\i}sica da Universidade do Porto, Rua das Estrelas, 4150-762 Porto, Portugal}
\affiliation{Instituto de Astrof\'{\i}sica e Ci\^encias do Espa\c co, Universidade do Porto, Rua das Estrelas, 4150-762 Porto, Portugal}
\author{L. Vacher}
\email{Leo.Vacher@etu.univ-grenoble-alpes.fr}
\affiliation{Universit\'e Grenoble-Alpes, 621 Avenue Centrale, 38400 Saint-Martin-d'H\`eres, France}

\date{8 November 2019}

\begin{abstract}
One of the clear predictions of string theory is the presence of a dynamical scalar partner of the spin-2 graviton, known as the dilaton. This will violate the Einstein Equivalence Principle, leading to a plethora of possibly observable consequences which is a cosmological context include dynamical dark energy and spacetime variations of nature's fundamental constants. The runaway dilaton scenario of Damour, Piazza and Veneziano is a particularly interesting class of string theory inspired models which can in principle reconcile a massless dilaton with experimental data. Here we use the latest background cosmology observations, astrophysical and laboratory tests of the stability of the fine-structure constant and local tests of the Weak Equivalence Principle to provide updated constraints on this scenario, under various simplifying assumptions. Overall we find consistency with the standard $\Lambda$CDM paradigm, and we improve the existing constraints on the coupling of the dilaton to baryonic matter by a factor of six, and to the dark sector by a factor of two. At the one sigma level the current data already excludes dark sector couplings of order unity, which would be their natural value.
\end{abstract}
\maketitle
\allowdisplaybreaks

%%%%%%%%%%%%%%%%%%%%%%%%%%%%%%%%%%%%%%%%%%%%%%%%%%%%%%%%%%%%%%%%%%%%%%%%%%

\section{Introduction}

The observational evidence for cosmic acceleration, first inferred from the luminosity distance of type Ia supernovae in 1998 \cite{SN1,SN2}, shows that our canonical theories of fundamental cosmology are incomplete (or possibly incorrect) and that there is new physics still to be discovered \cite{Accel1,Accel2}. String theory is arguably the most compelling available candidate to extend the current paradigm. One of its simplest predictions is the presence of a scalar partner of the spin-2 graviton, known as the dilaton (and hereafter denoted $\phi$). However, a massless dilaton would be in violent disagreement with experimental and observational constraints. The runaway dilaton scenario, introduced by Damour, Piazza and Veneziano \cite{DPV1,DPV2}, provides a conceptually appealing way to reconcile a massless dilaton with experimental data. Briefly speaking, the dilaton decouples while cosmologically attracted towards infinite bare coupling, and the coupling functions have a smooth finite limit
\begin{equation}\label{eq:coupfunc}
B_i(\phi)=c_i+{\cal O}(e^{-\phi})\,.
\end{equation}

In \cite{DPV2} the authors also noted that if the dilaton has an order unity coupling to the dark sector, the runaway of the dilaton towards strong coupling may yield dynamical dark energy, violations of the Equivalence Principle and variations of the fine-structure constant $\alpha$, all of which  that are potentially measurable. Conversely, if this coupling is much smaller the aforementioned effects are correspondingly smaller, and the model is effectively indistinguishable from the canonical $\Lambda$CDM. More recently, this analysis was updated in \cite{Dill1}, who provided the first quantitative constraints on this scenario, while the prospects for improving these constraints with the next generation of astrophysical facilities---specifically the Extremely Large Telescope and its high-resolution spectrograph ELT-HIRES \cite{EELT}---were discussed in \cite{Dill2}.

In recent years a number of stringent relevant data sets appeared, including background cosmology data and more sensitive astrophysical measurements of $\alpha$ and experimental tests of the Weak Equivalence Principle. The latter, from the MICROSCOPE satellite \cite{Touboul}, is particularly relevant, improving by an order of magnitude previous constrains from torsion balance tests and lunar laser ranging \cite{Torsion,Lunar}. It is therefore timely to revisit and update the constraints on these models. In what follows we therefore update the analysis of \cite{Dill1} in two different ways. Firstly, we use the latest available data. Secondly, while we start with the linearized approximation developed in \cite{DPV1,DPV2} and used in \cite{Dill1} (in which there is a simple analytic solution), we also obtain constraints for the full model (which have to be obtained by numerical integration), under several simplifying assumptions. In both cases we will see that the MICROSCOPE constraint helps to break a critical degeneracy between model parameters, thus leading to significantly improved constraints.

\section{Runaway dilaton cosmology}

As discussed in \cite{DPV1,DPV2}, the Einstein frame Lagrangian for the runaway dilaton scenario is
\begin{equation}\label{eq:lagr}
{\cal L}=\frac{R}{16\pi G}-\frac{1}{8\pi G}\left(\nabla\phi\right)^2-\frac{1}{4}B_F(\phi)F^2+... \,.
\end{equation}
where $R$ is the Ricci scalar and $B_F$ is the gauge coupling function. In this section we summarize the background cosmological evolution of these models (under the usual Friedmann-Lemaitre-Robertson-Walker assumptions) as well as its salient astrophysical and local signatures. In parallel we also introduce the various data sets that will be used (in the following section) to constrain the scenario, and describe the rest of our assumptions and statistical priors.

\subsection{Background cosmology}

One can show \cite{DPV2} that the Friedmann equation in this scenario is as follows
\begin{equation}\label{eq:friedmann}
3H^2=8\pi G\sum_i \rho_i+H^2\phi'^2\,,
\end{equation}
where the sum is over the standard components of the universe, including the dilaton potential but not its kinetic part which is described by the last term, and the prime is the derivative with respect to the logarithm of the scale factor. The total energy density and pressure of the field are the sum of the kinetic and potential parts
\begin{equation}\label{phidens}
\rho_\phi=\rho_k+\rho_v=\frac{(H\phi')^2}{8\pi G}+V(\phi)
\end{equation}
\begin{equation}\label{phipres}
p_\phi=p_k+p_v=\frac{(H\phi')^2}{8\pi G}-V(\phi)\,.
\end{equation}
The dilaton's contribution to the cosmological expansion is akin to that of a quintessence field through its potential (which we henceforth assume to be equivalent to a cosmological constant), though with an important difference: a modified Friedmann equation due to a kinetic term in which the usual $(d\phi/dt)^2$ dependence is multipled by a constant pre-factor which differs from the canonical quintessence one . This can also be seen by re-writing the Friedmann equation as
\begin{equation}\label{eq:ezfried}
\frac{H^2(z)}{H_0^2}=\frac{\Omega_m(1+z)^3+\Omega_v}{1-\phi'^2/3} \,,
\end{equation}
where $\Omega_m$ and $\Omega_V$ are the present-day values of the matter and dark energy densities. In what follows we will assume flat universes, so that $\Omega_m+\Omega_V=1-\phi_0'^2/3$, which can be used to replace $\Omega_V$ in the previous equation.

As background cosmology probes, we use the recent Pantheon catalogue of Type Ia supernovas of \citet{Riess} together with the compilation of 38 measurements of the Hubble parameter by \citet{Farooq}. The latter is useful for extending the redshift lever arm and thus increasing the overlap with the range of redshifts of $\alpha$ measurements discussed in the next subsection. Specifically, the Hubble parameter measurements reach up to redshift $z\sim2.36$, while the supernova data is at $z<1.5$ and the $\alpha$ measurements reach $z\sim4$. The Hubble constant is always analytically marginalized, following the prescription of \cite{Anagnostopoulos}.

From the above Lagrangian one also obtains the evolution equation for the dilaton
\begin{equation}\label{eq:field}
\frac{2}{3-\phi'^2}\phi''+\left(1-\frac{p}{\rho}\right)\phi'=-\sum_i\alpha_i(\phi)\frac{\rho_i-3p_i}{\rho}\,.
\end{equation}
Here $p=\sum_ip_i$, $\rho=\sum_i\rho_i$, and sums are again over all components except the kinetic part of the scalar field.

The $\alpha_i(\phi)$ are the couplings of the dilaton with each component $i$---baryons, dark matter and effective dark energy---which in principle are three independent parameters. In what follows we assume the coupling to the effective dark energy, $\alpha_V$, to be a constant, while the coupling to baryonic matter should be given by the logarithmic derivative of the QCD scale \cite{Polyakov}, which with the assumption that all gauge field couple, near the string cutoff, to the same gauge kinetic function, leads to
\begin{equation}\label{alphahadrel}
\frac{\alpha_{had}(\phi)}{\alpha_{had,0}}=e^{-(\phi-\phi_0)}\,,
\end{equation}
where $\phi_0$ is the value of the field today. As will become apparent, we can set $\phi_0=0$ without loss of generality---which we do in our numerical implementation. Following \cite{Dill1,Dill2} we consider three different possibilities for the dark matter coupling $\alpha_m$:
\begin{itemize}
\item Dark Coupling: $\alpha_m=\alpha_V$
\item Matter Coupling: $\alpha_m(\phi)=\alpha_{had}(\phi)$
\item Field Coupling : $\alpha_m(\phi)=-\phi'$
\end{itemize}
These three phenomenological choices, all motivated by the discussion in \cite{DPV2}, span the range of behaviours in these models. The dark coupling corresponds to the simplest assumption that the dark sector is characterized by as single coupling, applicable to both dark matter and dark energy. Conversely the matter coupling assumes a single coupling for dark matter and baryons, while dark energy couples differently. Finally the field coupling corresponds to the approximate matter-era solution discussed in  \cite{DPV2}. In all three cases the free parameters are therefore $\alpha_{had0}$ and $\alpha_V$.

Experimental constraints impose a tiny coupling to baryonic matter, discussed in the following subsection. In these models, this may naturally emerge due to a Damour-Polyakov type screening of the dilaton \cite{Polyakov}. On the other hand, dark matter and dark energy couplings of order unity are allowed by experimental constraints \cite{Torsion}, but constrained to be less than about 0.20 when combining local and astrophysical constraints \cite{Dill1}.

\subsection{Local and astrophysical constraints}

There are several additional constraints that can be added to our analysis. Firstly the Eddington parameter $\gamma$, which quantifies the amount of deflection of light by a gravitational source, has the value
\begin{equation}\label{eddingt}
\gamma-1=-2\alpha_{had,0}^2\,,
\end{equation}
and is constrained by the Cassini bound, $\gamma-1=(2.1\pm2.3)\times10^{-5}$ \cite{Cassini}. However, this is totally subdominant when compared to the constraint on the E\"{o}tv\"{o}s parameter, quantifying violations to the Weak Equivalence Principle, which in these models is
\begin{equation}\label{eotvos}
\eta\sim5.2\times10^{-5}\alpha_{had,0}^2\,.
\end{equation}
The recent MICROSCOPE satellite result constrains this to be $\eta=(-0.1\pm1.3)\times10^{-14}$, so we approximately have $|\alpha_{had,0}|<10^{-4}$; as we will see this can be significantly improved by combination with the rest of the available data sets.

We emphasize that the constraints in the above paragraph do not apply to the dark sector, whose couplings may be close to unity. In the runaway dilaton scenario there are two broad cases to consider. The first is that the dark sector couplings are also much smaller than unity; if so the small field velocity leads to violations of the Equivalence Principle that are effectively undetectable. In this case the contributions of the kinetic and potential parts of the scalar field will be $\Omega_k={\phi'}^2/3\ll1$ and $\Omega_V\sim0.7$. The second (and more interesting) case is that the dark couplings saturate their bounds and are (close to) order unity. If so, violations of the Equivalence Principle and variations of the fine-structure constant (to be discussed presently) are typically larger. In this case $\Omega_k$ may be more significant, and $\Omega_V$ correspondingly smaller \cite{quint}.

The present value of the field derivative is also constrained: during matter-domination the equation of state has the form
\begin{equation}
w_m(\phi)=\frac{1}{3} {\phi'}^2\,.
\end{equation}
Recent analyses \cite{Matter1,Matter2} constrain the matter equation of state to the (conservative) value $|w_m|<0.003$, which we translate to the following prior on the field derivative today (an initial condition for integrating the dilation equation): $|\phi_0'|=0.0\pm0.1$.

Moreover, these models necessarily lead to space-time variations of the fine-structure constant $\alpha$, which are tightly constrained---for a recent review see \cite{ROPP}. Consistently with our previous assumption that all gauge fields couple to the same gauge kinetic function, here $\alpha$ will be proportional to $B_F^{-1}(\phi)$. One then finds that \cite{Dill1}
\begin{equation}\label{evolfull}
\frac{\Delta\alpha}{\alpha}(z)\equiv\frac{\alpha(z)-\alpha_0}{\alpha_0}=B_F^{-1}(\phi(z))-1 =\frac{\alpha_{had,0}}{40}\left[1-e^{-(\phi(z)-\phi_0)}\right]\,.
\end{equation}
Thus the evolution of $\alpha$ depends both on the baryonic matter coupling (which provides an overall normalization) and on the speed of the field. This evolution is therefore constrained by  astrophysical (spectroscopic) tests of the stability of $\alpha$. We will use both the Webb {\it et al.} \cite{Webb} data set (a large data set of 293 archival data measurements) and a smaller but more recent data set of 24 dedicated measurements \cite{ROPP,Cooksey}. The former spans a redshift range $0.22\le z \le 4.18$, while the latter spans a narrower range, $1.02\le z \le 2.13$ but contains more stringent measurements which are expected to have a better control of possible systematics. All this data comes from high-resolution spectroscopy comparisons of optical/UV fine-structure atomic doublets, which are only sensitive to the value of $\alpha$---and not, say, to the values of particle masses, ratios of which can be probed by other means \cite{Kozlov}.

Additionally we use the geophysical constraint from the Oklo natural nuclear reactor \cite{Oklo}, $\Delta\alpha/\alpha =(0.005\pm0.061)\times10^{-6}$, at an effective redshift $z=0.14$ and under the simplifying assumption that $\alpha$ is the only parameter than may have been different and all the remaining physics is unchanged.

Last but not least, the time drift of $\alpha$ is given by
\begin{equation}\label{alphazero}
\frac{1}{H}\frac{\dot\alpha}{\alpha}=\frac{\alpha_{had}}{40}{\phi'}e^{-(\phi(z)-\phi_0)} \,.
\end{equation}
In particular this equation applies at the present day (describing the current running of $\alpha$) and this variation is constrained by the atomic clocks bound \cite{Rosenband}
\begin{equation} \label{rosen}
\left(\frac{1}{H}\frac{\dot\alpha}{\alpha}\right)_0=(-0.22\pm0.32)\times10^{-6}\,.
\end{equation}
Thus local atomic clock experiments constrain the product of the baryonic coupling and the field speed today.

\section{Current constraints}

We are now ready to constrain the runaway dilaton scenario using the data sets introduced in the previous section. We will start by doing this using a linearized approximation (effectively corresponding to a slow-roll approximation), originally discussed in \cite{DPV1,DPV2} and also used in \cite{Dill1}. We then move on to study the general case, under one of the three previously simplifying assumptions (dark coupling, matter coupling or field coupling). We will find that all three assumptions lead to very similar constraints, which are also comparable to those of the slow roll approximation.

\subsection{Linearized approximation}

Since in the current work we will be interested in the observational consequences of the model at low redshifts, and given the expectation that the field dynamics should be relatively slow, one could think of linearizing the field evolution
\begin{equation}\label{evollinear}
\phi\sim\phi_0+{\phi_0'}\ln{a}\,,
\end{equation}
in which case the Friedmann equation becomes (with the additional assumption of a flat universe) simplifies to
\begin{equation}\label{eq:ezfriedlin}
\frac{H^2(z)}{H_0^2}=1+\Omega_m\frac{(1+z)^3-1}{1-\phi_0'^2/3} \,,
\end{equation}
while the redshift dependence of the fine-structure constant takes the form
\begin{equation}\label{evolslow}
\frac{\Delta\alpha}{\alpha}(z)\approx\, -\frac{1}{40}\alpha_{had,0} {\phi_0'}\ln{(1+z)}\,.
\end{equation}
Note that this approximate solution does not depend on the two dark sector couplings.

%%%%%%%%%%%%%%%%%%%%%%%%%%%%%%%%%%%%%%%
\begin{figure}
\includegraphics[width=3in,keepaspectratio]{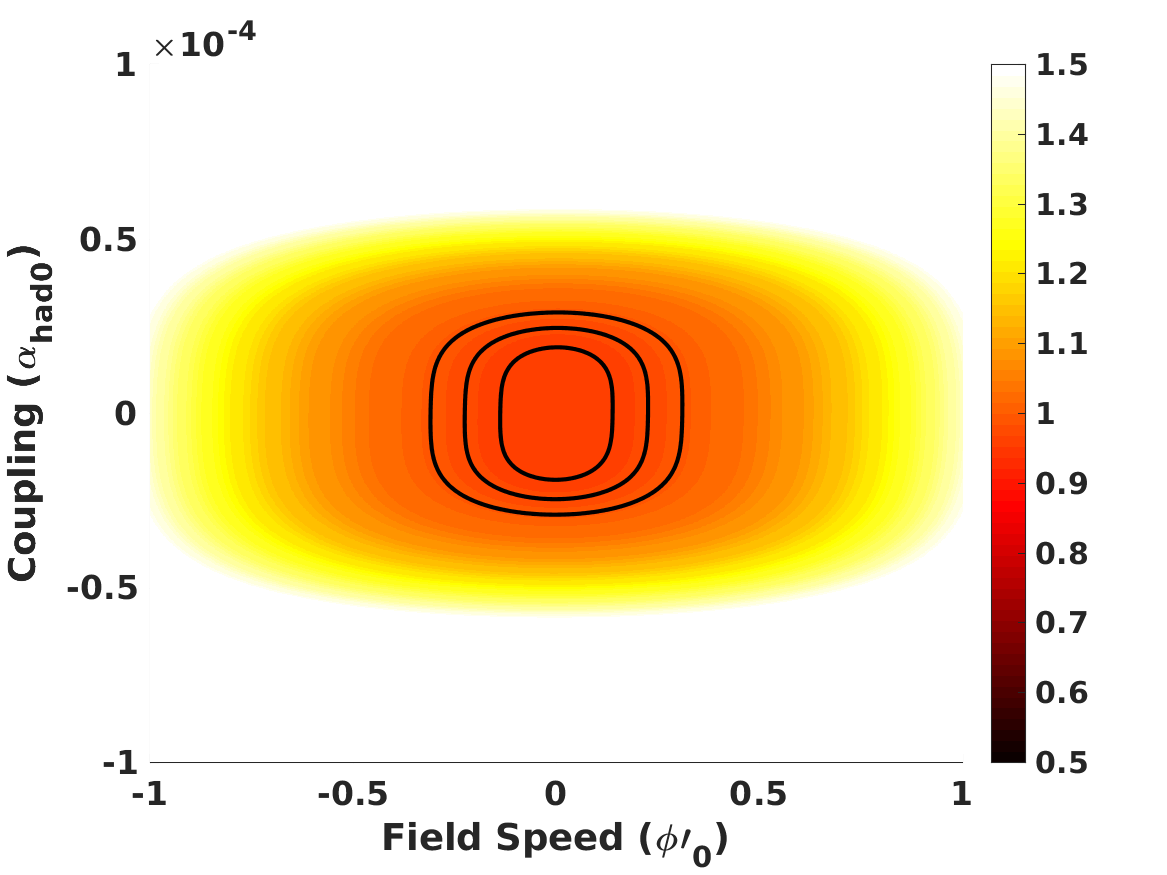}
\includegraphics[width=3in,keepaspectratio]{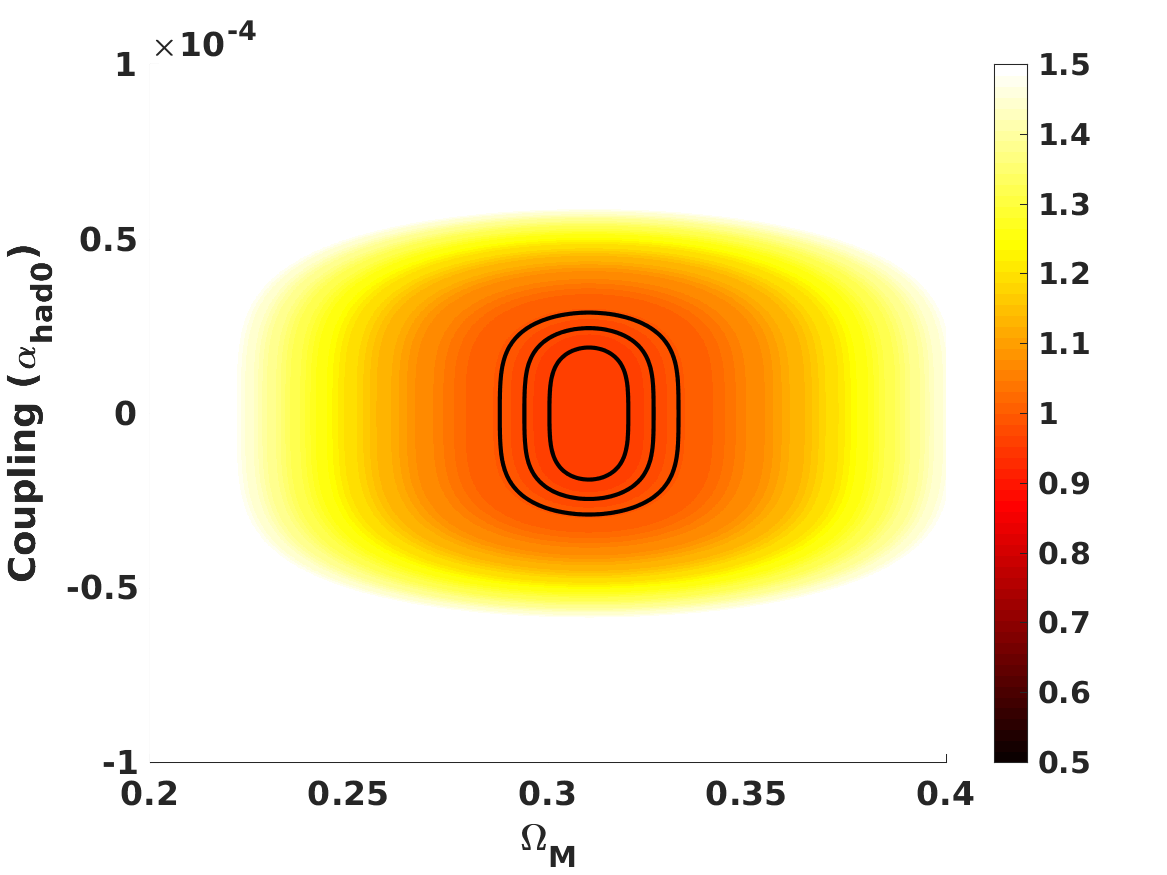}
\includegraphics[width=3in,keepaspectratio]{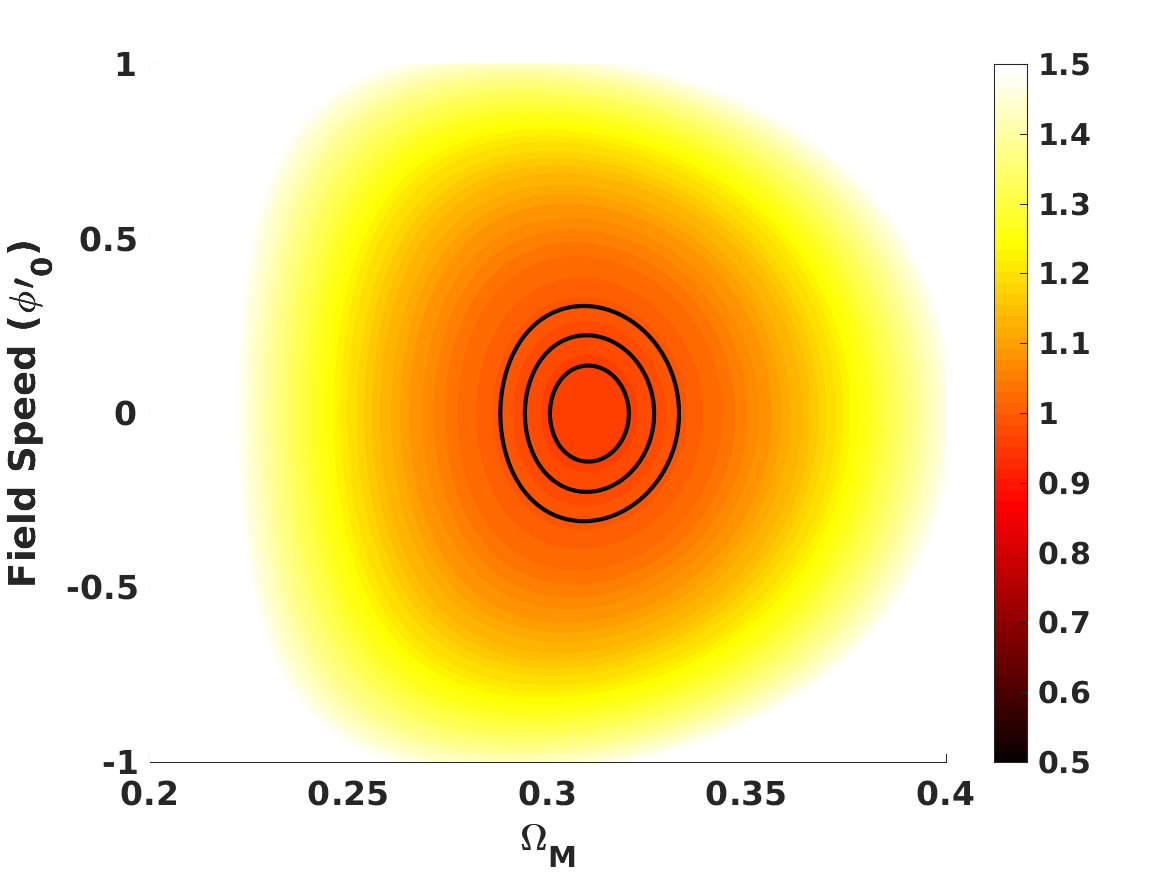}
\caption{Likelihood constraints in the various 2D planes for the runaway dilaton scenario in the linearized approximation. The black lines represent the one, two and three sigma confidence levels, and the colormap depicts the reduced chi-square of the fit. The value of the reduced chi-square at the best fit is $\chi^2_\nu=0.99$.}
\label{fig1}
\end{figure}
%%%%%%%%%%%%%%%%%%%%%%%%%%%%%%%%%%%%%%%

We carry out a standard statistical analysis assuming uniform priors on the present-day values of the baryonic coupling and field speed, while for the matter density we use a Planck-like prior, $\Omega_m=0.315\pm0.007$ \cite{Planck}. The results of this analysis are summarized in Fig. \ref{fig1}. As expected there are no significant degeneracies between the model parameters, since each of them is separately constrained by a different observable: MICROSCOPE for the baryonic coupling, the matter equation of state for the field speed, and Planck for the matter density. The posterior one-sigma confidence levels are
\begin{equation}
\alpha_{had,0}=(0.0\pm15.2)\times10^{-6}
\end{equation}
\begin{equation}
\phi_0'=0.00\pm0.09
\end{equation}
\begin{equation}
\Omega_m=0.310\pm0.006\,.
\end{equation}

%%%%%%%%%%%%%%%%%%%%%%%%%%%%%%%%%%%%%%%
\begin{figure}
\includegraphics[width=3in,keepaspectratio]{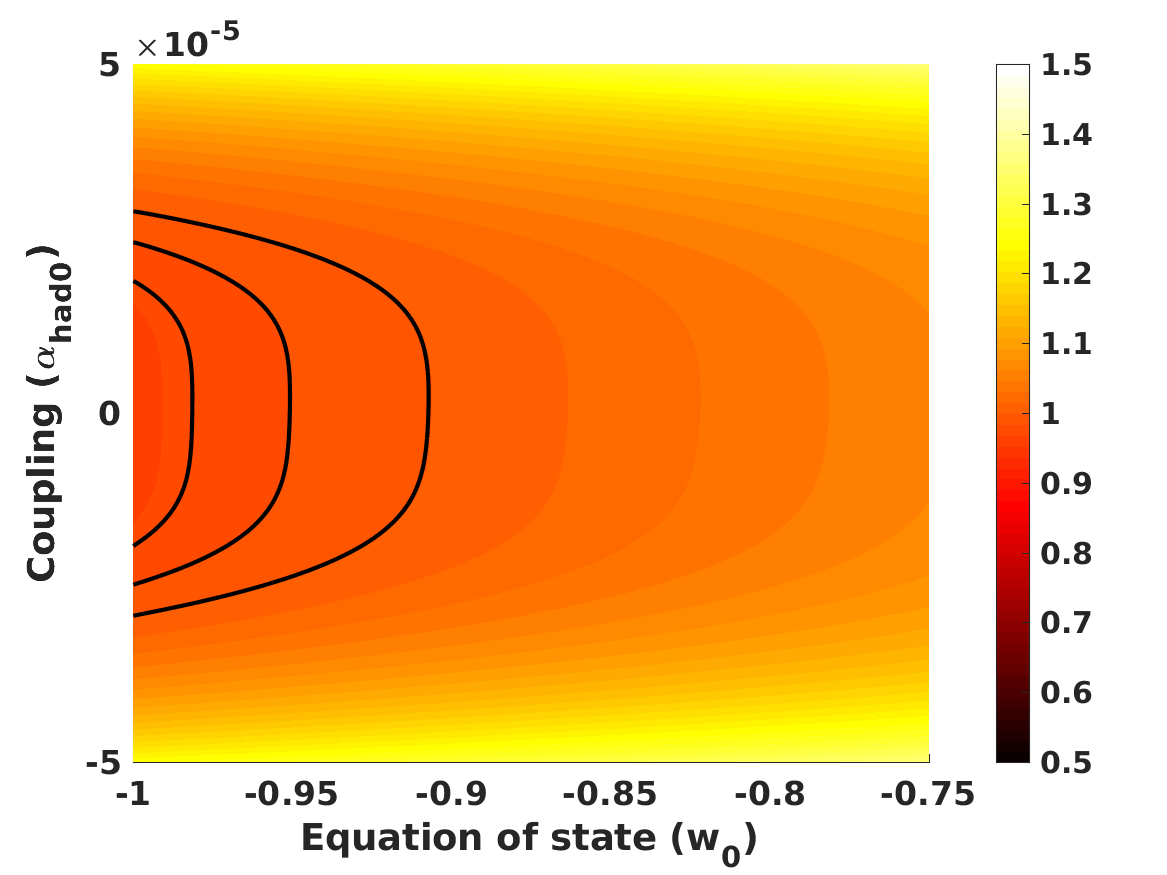}
\includegraphics[width=3in,keepaspectratio]{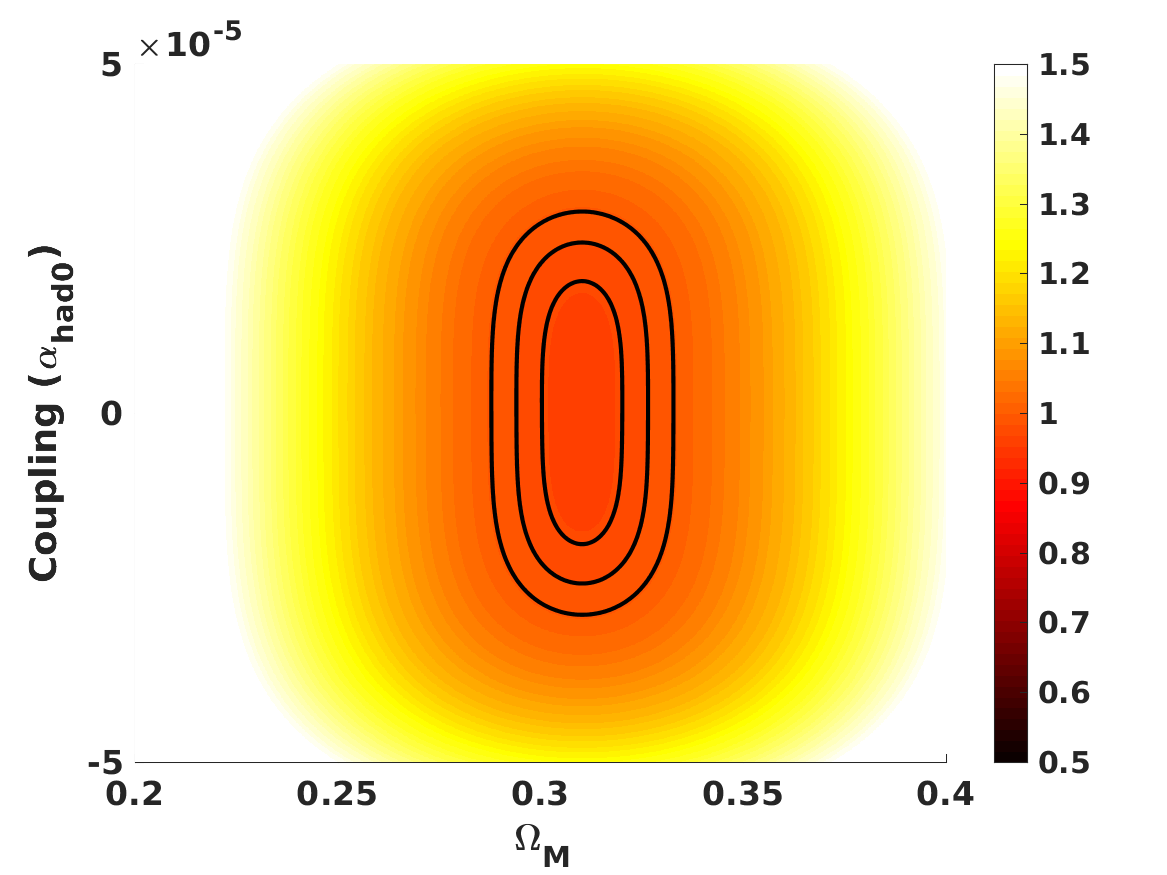}
\includegraphics[width=3in,keepaspectratio]{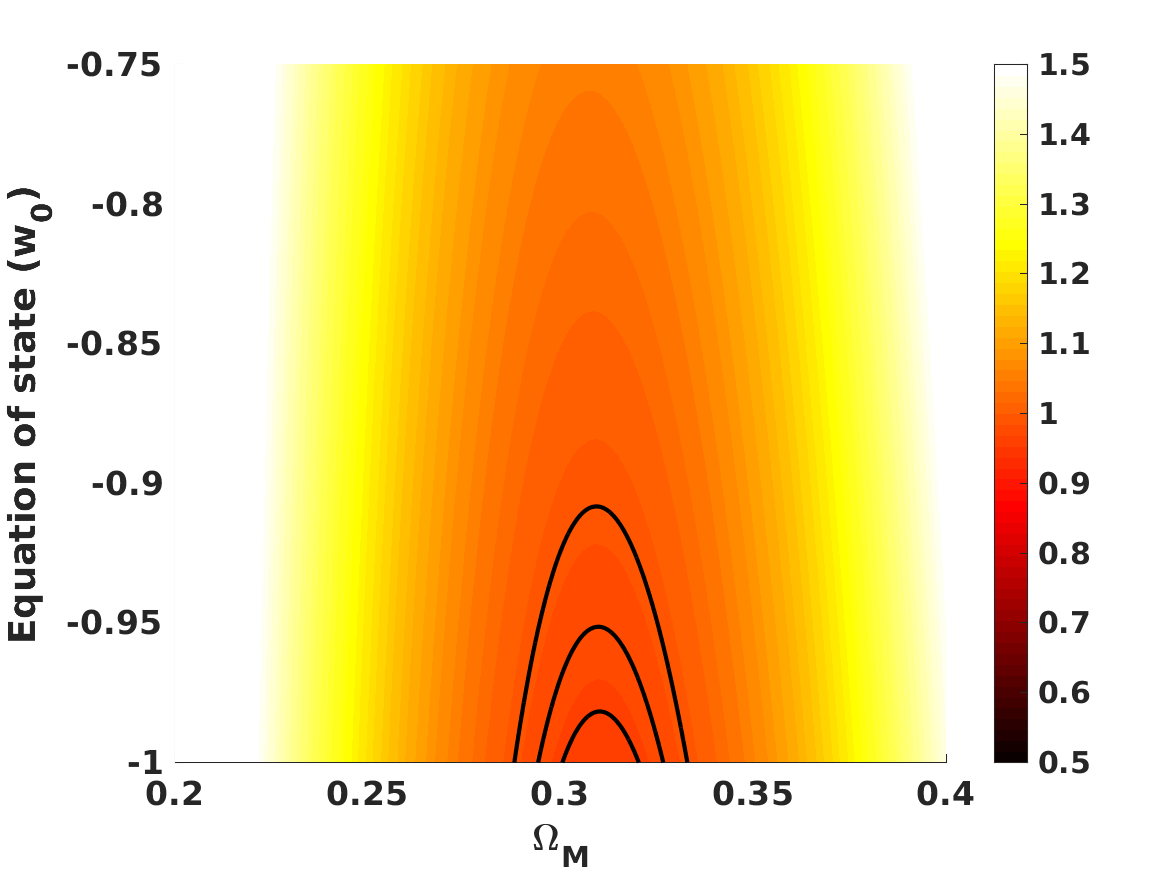}
\caption{Same as Fig. \protect\ref{fig1}, for the alternative parametrization including the effective dark energy equation of state. The value of the reduced chi-square at the best fit is $\chi^2_\nu=0.99$.}
\label{fig2}
\end{figure}
%%%%%%%%%%%%%%%%%%%%%%%%%%%%%%%%%%%%%%%

It is worthy of note that in this model the present-day value of the dark energy equation of state is
\begin{equation}\label{darkenergyeos}
1+w_0=\frac{2\Omega_{k}}{\Omega_{k}+\Omega_{v}}=\frac{2{\phi_0'}^2}{{\phi_0'}^2+3\Omega_V}\,;
\end{equation}
we can use this relation to replace the value of the field speed, and repeat the previous analysis. We assume a flat prior on $w_0$ but allow only canonical values, $w_0\ge-1$. The results of this analysis are depicted in Fig. \ref{fig2}. At the one-sigma confidence level we now find
\begin{equation}
\alpha_{had,0}=(1.8^{+13.4}_{-16.6})\times10^{-6}
\end{equation}
\begin{equation}
w_0<-0.992
\end{equation}
\begin{equation}
\Omega_m=0.310^{+0.006}_{-0.007}\,.
\end{equation}
The results are compatible with but not exactly the same as those of the previous case, the reason being that a uniform prior on the field speed is not equivalent to a uniform prior on the dark energy equation of state. In either case, the constraint on the baryonic coupling is improved by about a factor of six.

\subsection{Full model}

We now discuss constraints on the full model, in three separate cases: the dark coupling, matter coupling and field coupling assumptions introduced in the previous section. In addition to simplifying the analysis by reducing the number of free model parameters, they also serve as a way of exploring the range of allowed consequences of these models. As a further simplification, and given that we don't expect the matter density to strongly correlate with the dilaton coupling parameters, we assume fixed values of the baryon, dark matter and dark energy densities; specifically we have chosen values in agreement with the latest Planck results \cite{Planck}.

%%%%%%%%%%%%%%%%%%%%%%%%%%%%%%%%%%%%%%%
\begin{figure*}
\includegraphics[width=5.5cm,keepaspectratio]{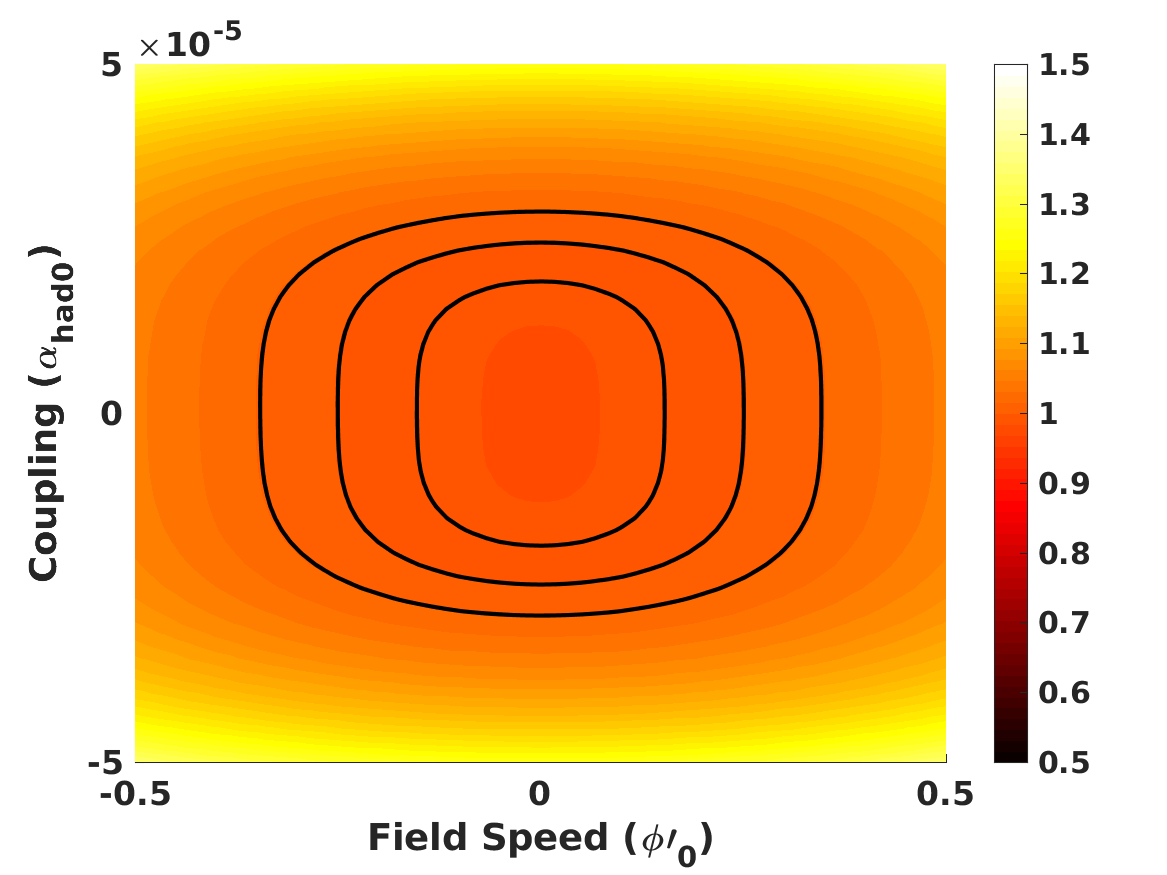}
\includegraphics[width=5.5cm,keepaspectratio]{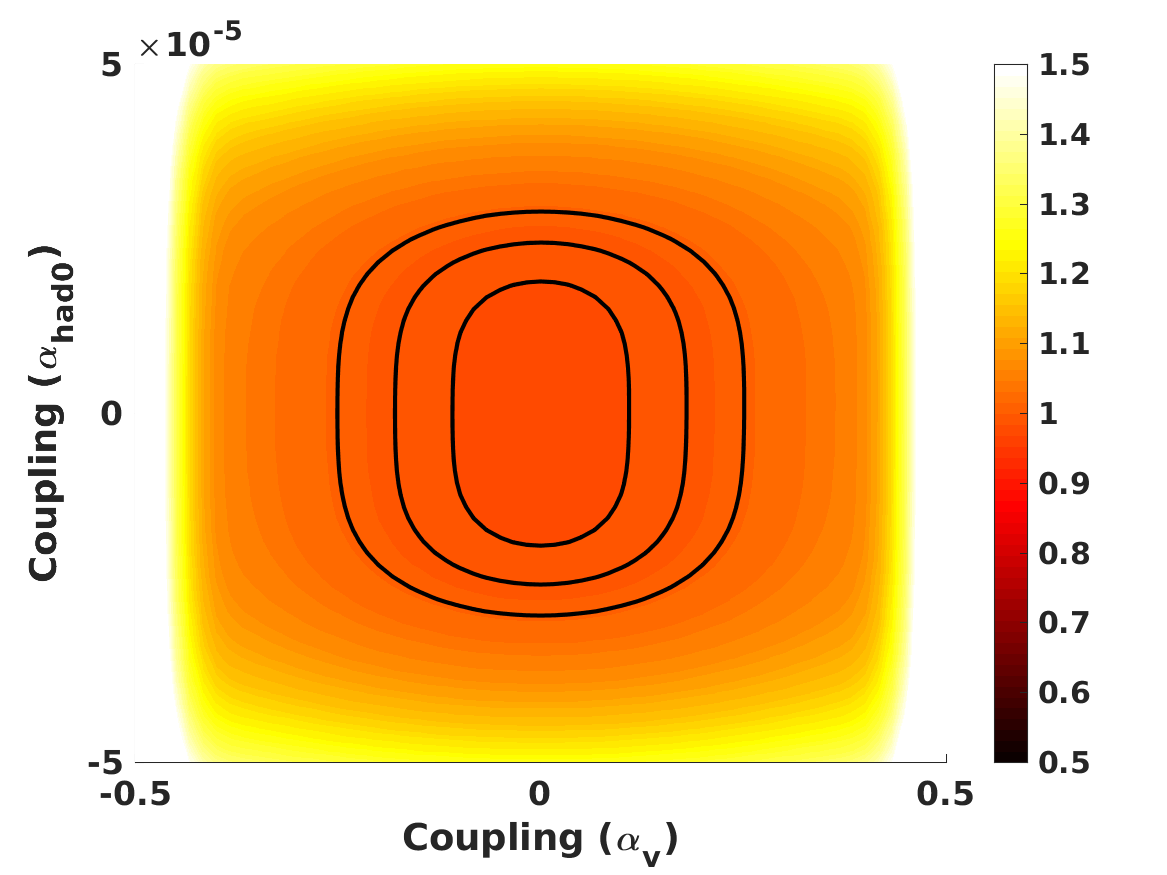}
\includegraphics[width=5.5cm,keepaspectratio]{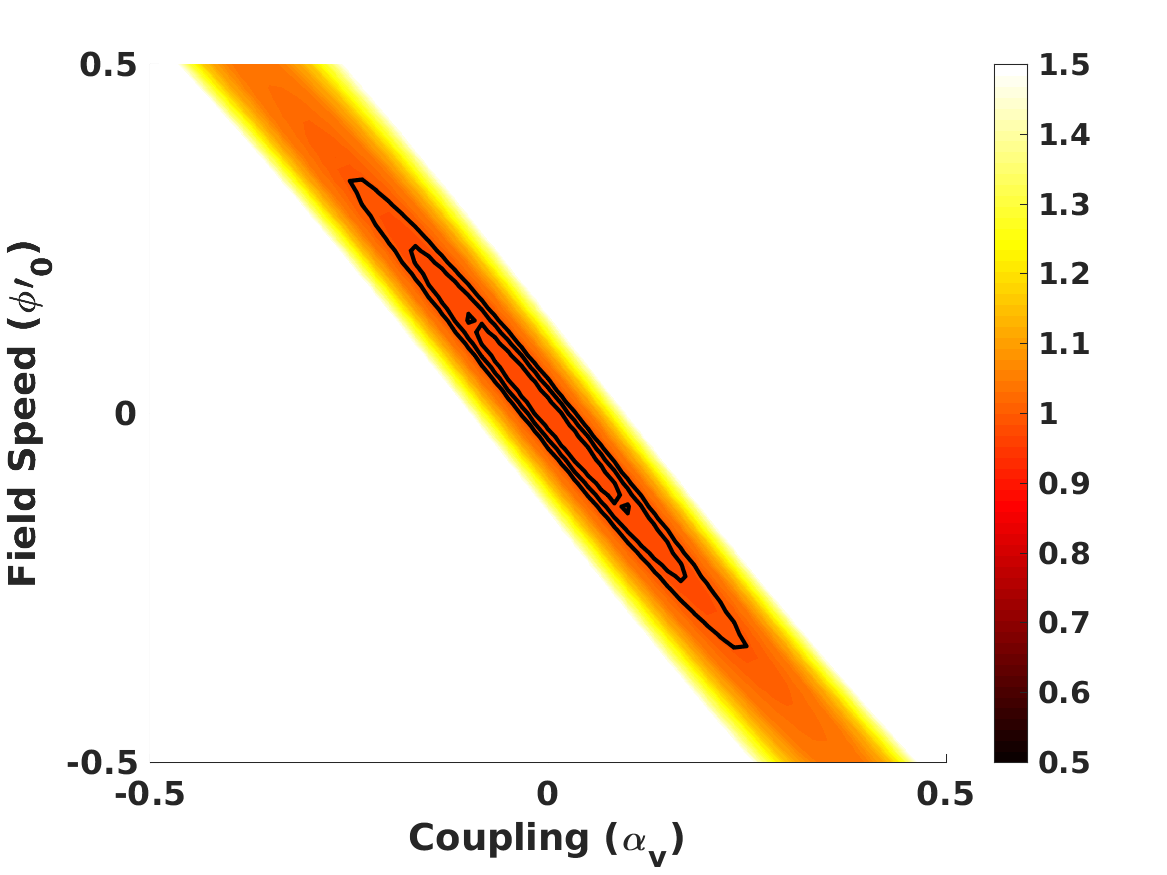}\\
\includegraphics[width=5.5cm,keepaspectratio]{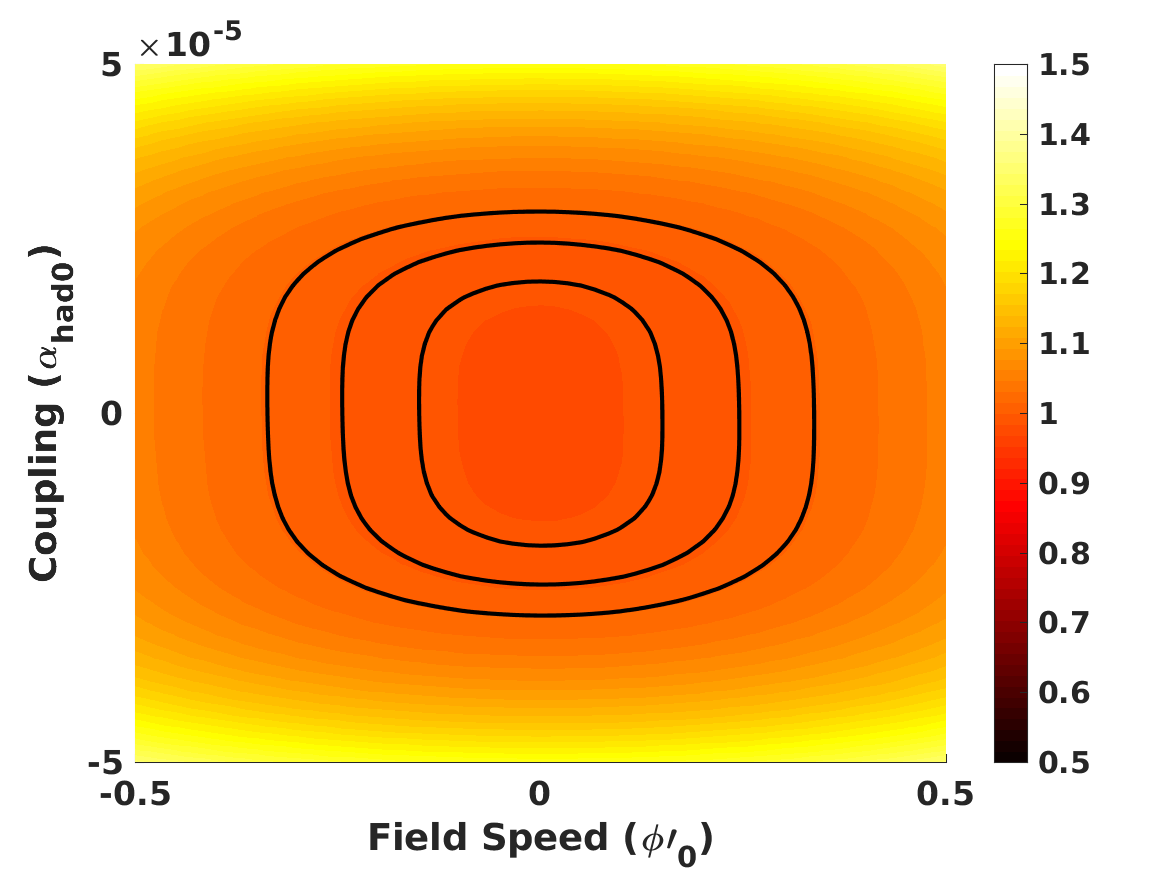}
\includegraphics[width=5.5cm,keepaspectratio]{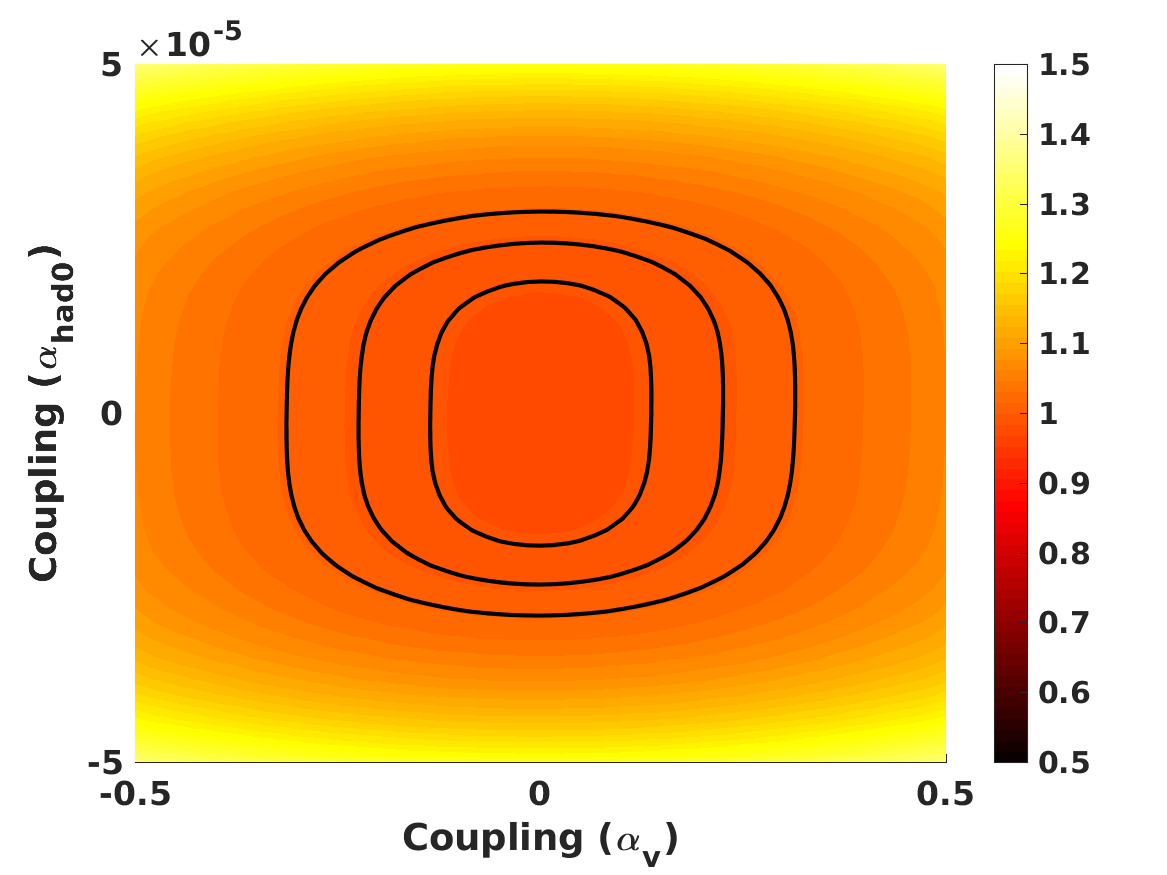}
\includegraphics[width=5.5cm,keepaspectratio]{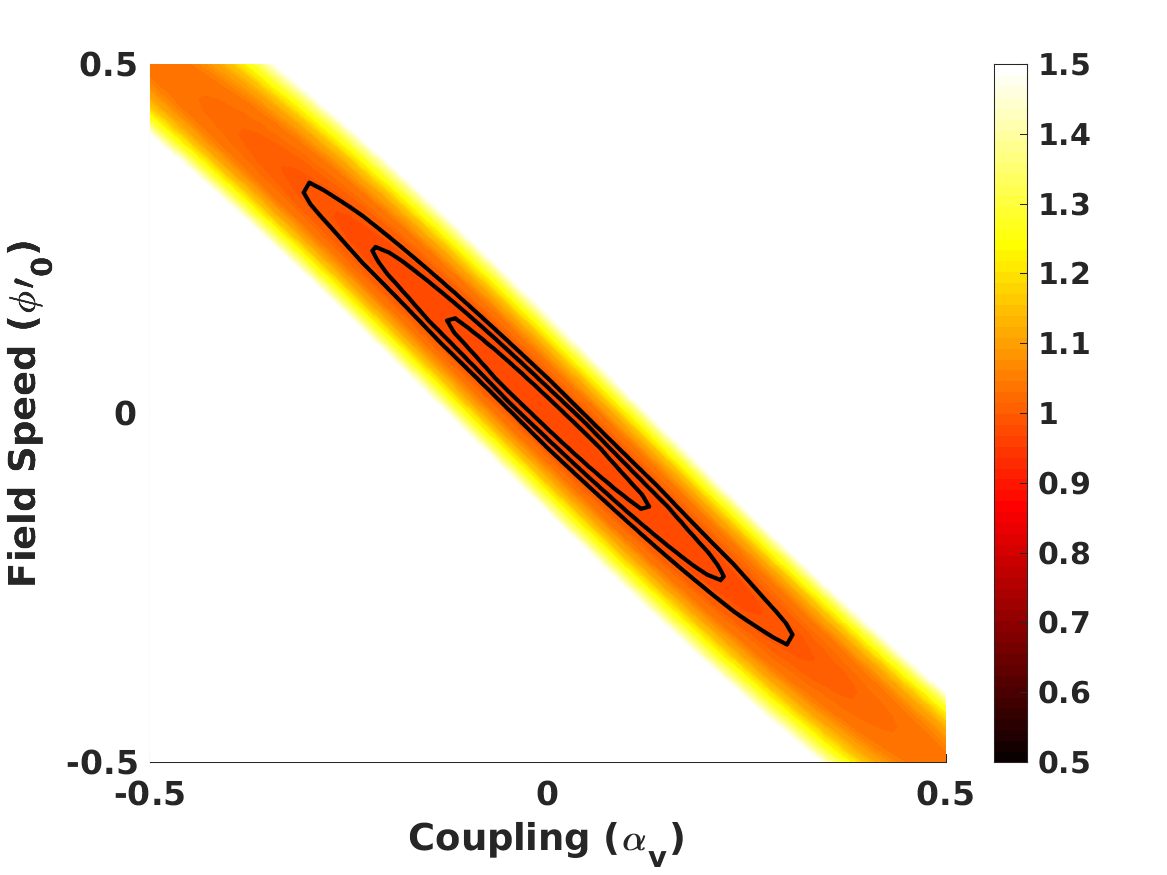}\\
\includegraphics[width=5.5cm,keepaspectratio]{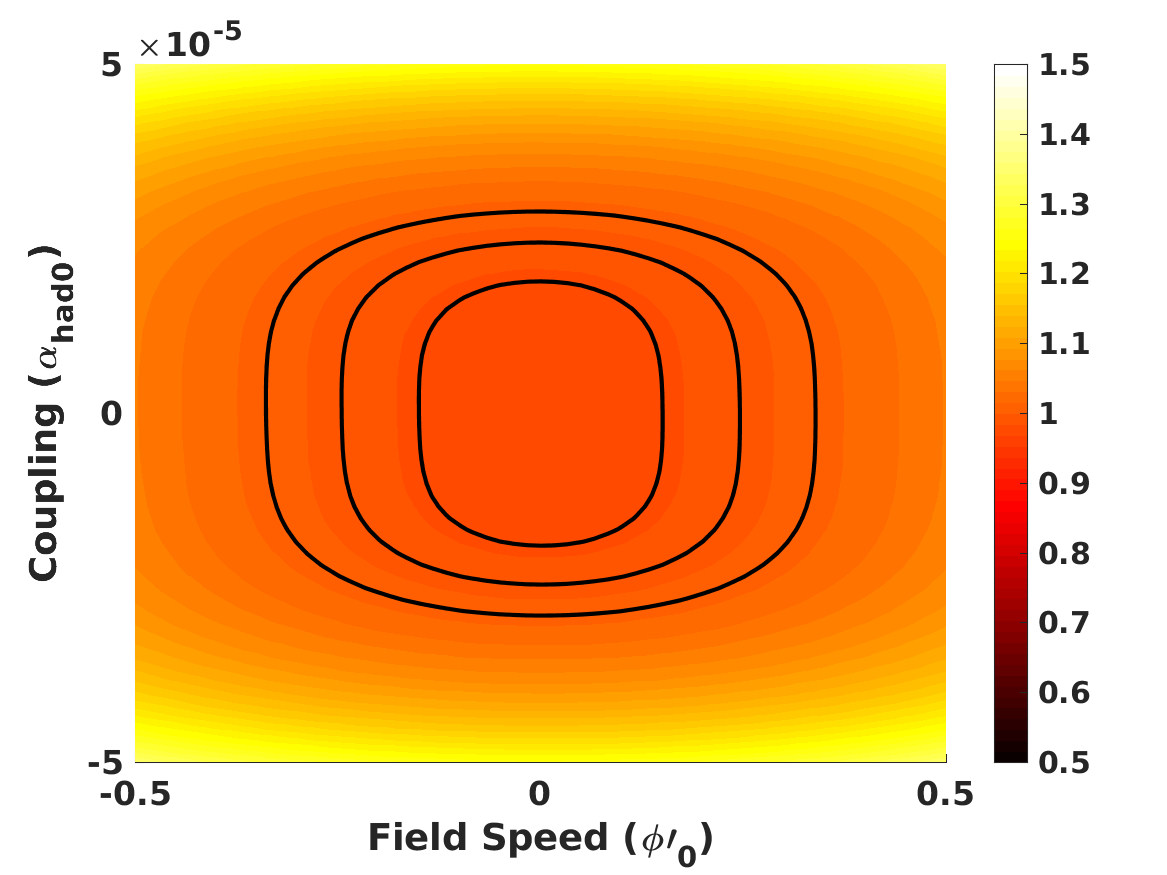}
\includegraphics[width=5.5cm,keepaspectratio]{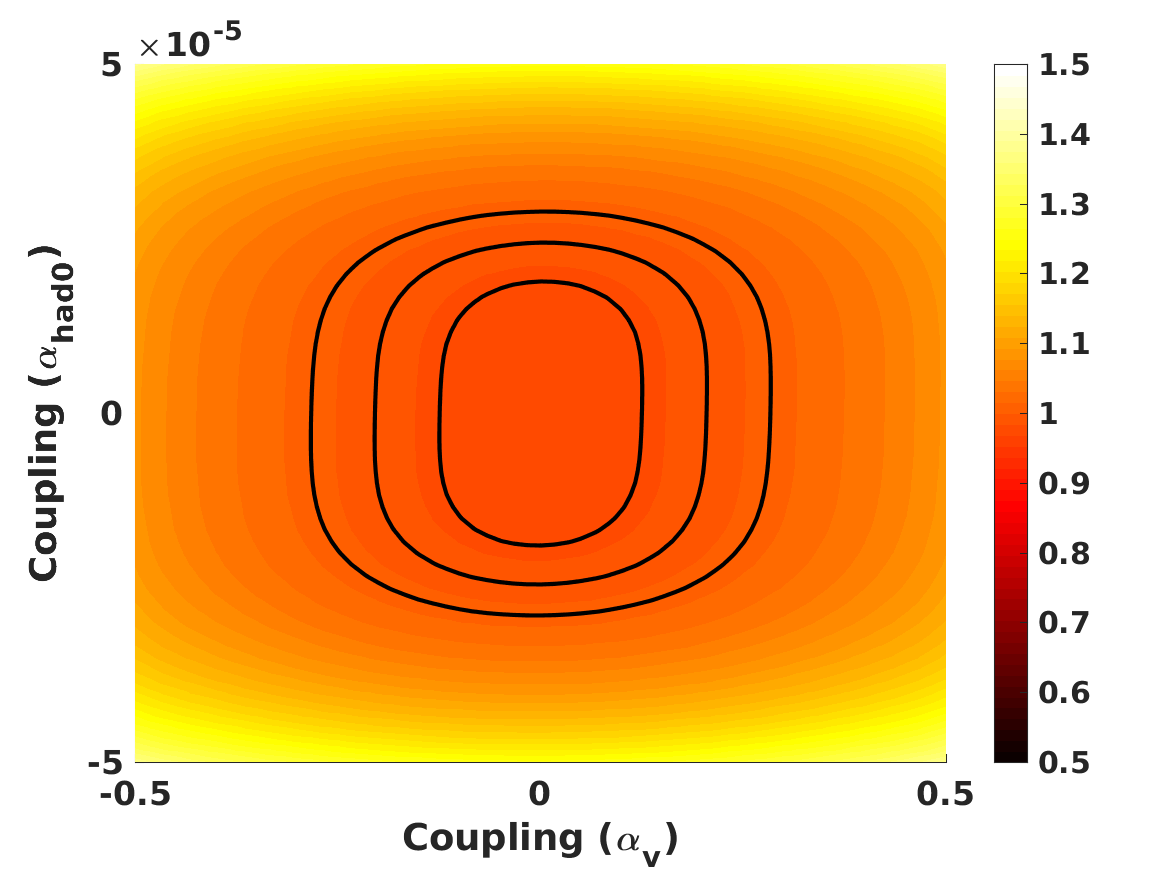}
\includegraphics[width=5.5cm,keepaspectratio]{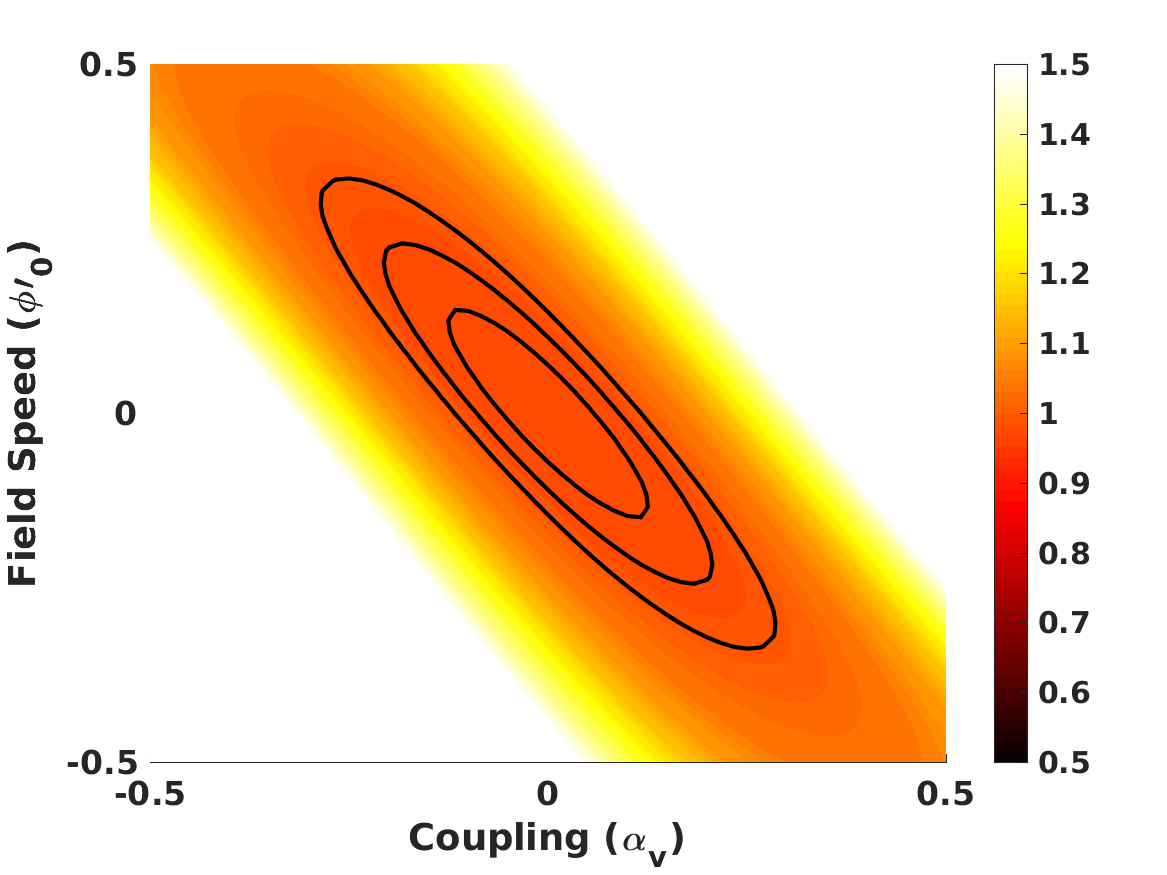}
\caption{Likelihood constraints in the various 2D planes for the runaway dilaton scenario. Top, middle and bottom plots respectively correspond to the dark coupling, matter coupling and field coupling assumptions. The black lines represent the one, two and three sigma confidence levels, and the colormaps depict the reduced chi-square of the fits. The value of the reduced chi-square at the best fit is $\chi^2_\nu=0.99$ in all three cases.}
\label{fig3}
\end{figure*}
%%%%%%%%%%%%%%%%%%%%%%%%%%%%%%%%%%%%%%%

\begin{figure}
\includegraphics[width=3in,keepaspectratio]{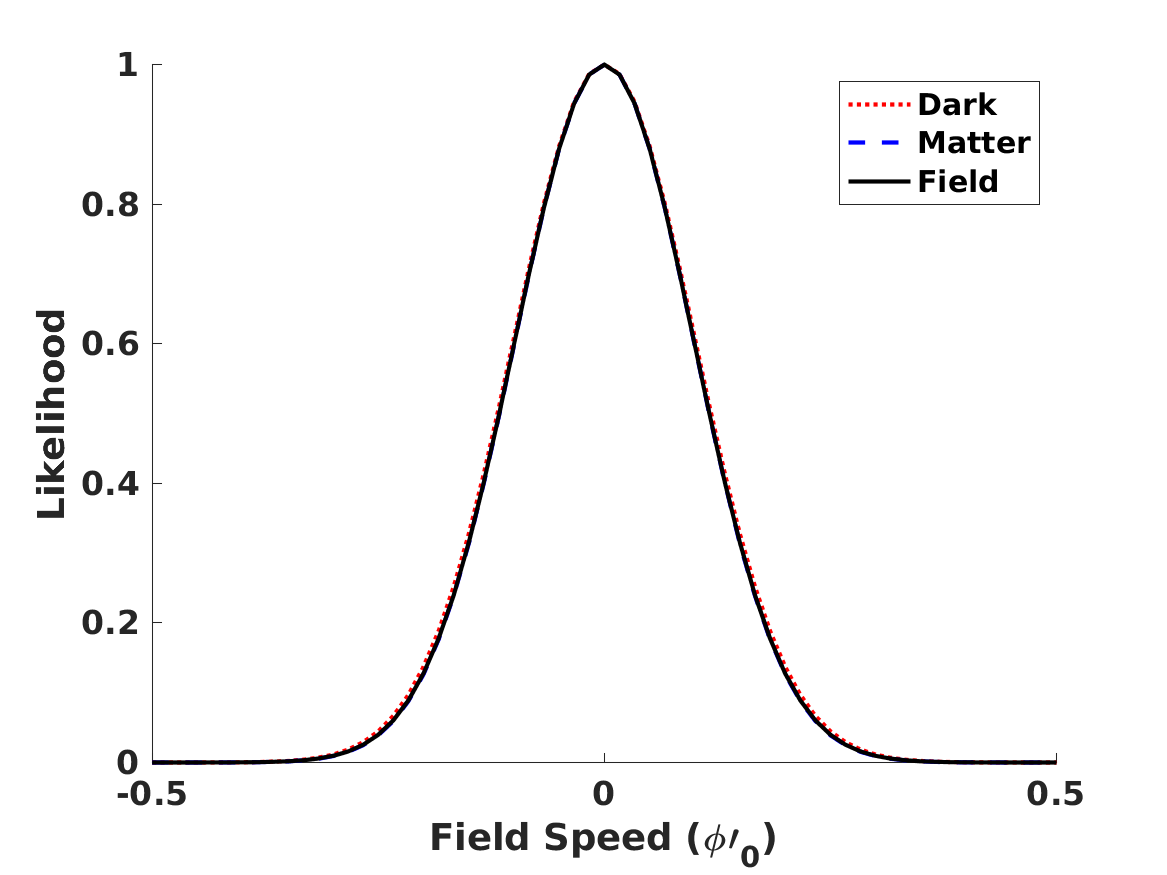}
\includegraphics[width=3in,keepaspectratio]{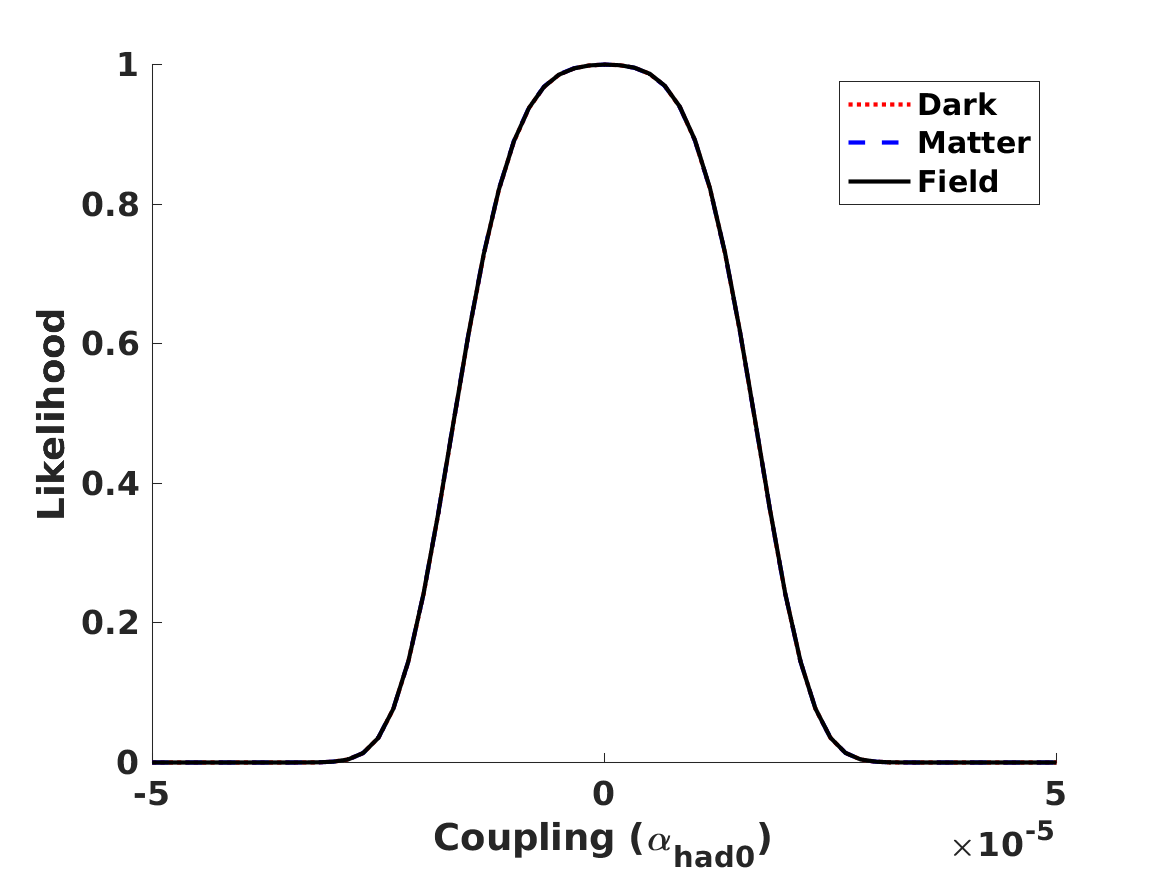}
\includegraphics[width=3in,keepaspectratio]{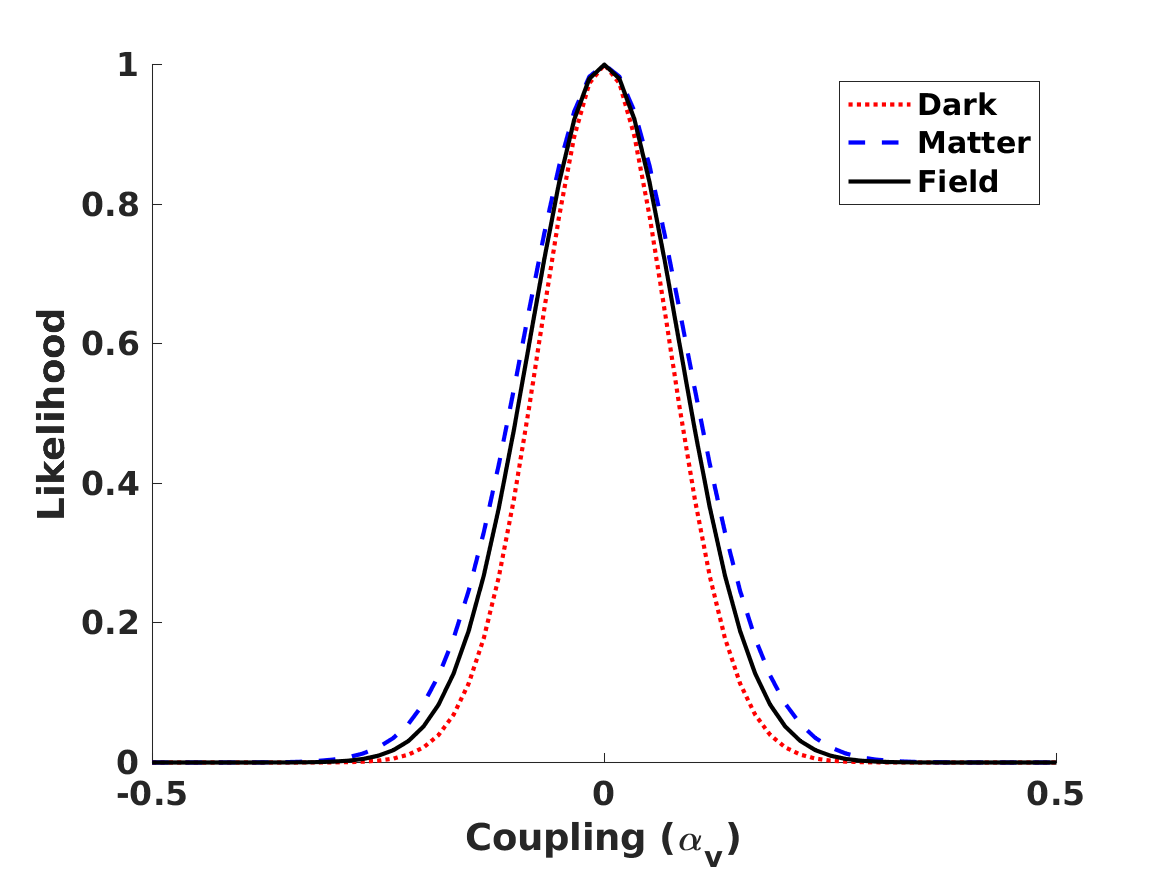}
\caption{One-dimensional posterior likelihoods for the full analysis in Fig. \protect\ref{fig3}. Comparing the results of the dark coupling, matter coupling and field coupling assumptions, the only significant differences occur for the dark energy coupling, $\alpha_V$.}
\label{fig4}
\end{figure}

The results of this analysis are presented in Fig. \ref{fig3} (which containing the relevant 2D likelihoods) and Fig. \ref{fig4} (with the corresponding 1D posterior likelihoods). We find that the results are quite similar in the three cases. The only degeneracy is between the dilaton couplings to baryons and dark energy, and the degeneracy direction is the only significant difference between the three scenarios. The one-sigma constraints on the field speed and baryonic coupling are almost exactly the same in all three cases
\begin{equation}
\alpha_{had,0}=(0\pm15)\times10^{-6}
\end{equation}
\begin{equation}
\phi_0'=0.0\pm0.1\,;
\end{equation}
the latter is directly coming from the assumed prior, while the former is again a factor of six improvement over previous constraints, and is mainly due to the MICROSCOPE measurement.

%%%%%%%%%%%%%%%%%%%%%%%%%%%%%%%%%%%%%%%
%%%%%%%%%%%%%%%%%%%%%%%%%%%%%%%%%%%%%%%

On the other hand, the one-sigma constraints on the dark energy coupling do vary in the three cases
\begin{equation}
\textrm{Dark coupling:}\,\quad \alpha_V=0.00\pm0.07
\end{equation}
\begin{equation}
\textrm{Matter coupling:}\,\quad \alpha_V=0.00\pm0.09
\end{equation}
\begin{equation}
\textrm{Field coupling:}\,\quad \alpha_V=0.00\pm0.08\,.
\end{equation}
These constraints improve those of \cite{Dill1} by a factor of two. In all cases, the result is fully consistent with $\Lambda$CDM, and show that these dark sector couplings are no longer allowed to be of order unity.

\section{Conclusion}

In this letter we took advantage of the recent improvements in sensitivity of background cosmology and high-resolution astrophysical spectroscopy data sets, as well as of local laboratory experiments, to improve constraints on the string theory inspired runaway dilaton scenario. Our results show consistency with the standard $\Lambda$CDM paradigm, but improve the existing constraints \cite{Dill1} on the coupling of the dilaton to baryonic matter by a factor of six, and to the dark sector by a factor of two. We have also confirmed that the previously used linearized (slow-roll) solution \cite{DPV2,Dill1} is a reasonable approximation for the purpose of constraining the baryonic coupling, although it is insensitive to the dark sector couplings. Finally we have also shown that different simplifying assumptions on the behaviour of the dark matter coupling lead to fairly similar results.

Our results go some of the way towards the improved constraints foreseen by \cite{Dill2} for next-generation astrophysical facilities. The main source of our reported improvement is the stringent MICROSCOPE constraint on the Weak Equivalence Principle \cite{Touboul}. Additional improvements are therefore expected from the final MICROSCOPE results (due to appear in the near future), and in the longer term with the planned STEP mission \cite{step}. Our work highlights how a combination of cosmological, astrophysical and local measurements can significantly constrain fundamental physical paradigms. In the case at hand, current data already excludes dark sector couplings of order unity, which would be their natural value. This shows that any allowed runaway dilation scenario will be, in terms of background cosmology, very similar to (and thus difficult to distinguish from) the standard $\Lambda$CDM paradigm. Thus the most likely observational route to identify cosmological signatures of these models would be through the detection of spacetime variations of the fine-structure constant. it remains to be seen whether this result for the runaway dilaton applies more widely to the string theory paradigm.

\begin{acknowledgments}

This work was financed by FEDER---Fundo Europeu de Desenvolvimento Regional funds through the COMPETE 2020---Operacional Programme for Competitiveness and Internationalisation (POCI), and by Portuguese funds through FCT---Funda\c c\~ao para a Ci\^encia e a Tecnologia in the framework of the project POCI-01-0145-FEDER-028987.

\end{acknowledgments}
\bibliography{dilaton}
\end{document}